\documentclass[10pt]{article}
\usepackage{aaspp4}
\usepackage{graphicx}
\usepackage{amstext}

\newcommand{\B}{{\mathbf B}}
\newcommand{\ve}{{\mathbf v}}
\newcommand{\E}{{\mathcal E}}
\newcommand{\lfrac}[2]{{{#1}{/}{#2}}}
\newcommand{\software}{\sc}

\begin{document}

\lefthead{MAGNETO-CENTRIFUGAL LAUNCHING OF JETS: I}
\righthead{{KRASNOPOLSKY}, {LI} AND {BLANDFORD}}

\title{Magneto-Centrifugal Launching of Jets from Accretion
Disks\\I: Cold Axisymmetric Flows}

\author{Ruben Krasnopolsky}
\affil{Theoretical Astrophysics, California Institute of Technology,
130--33 Caltech, Pasadena, CA 91125;
ruben@tapir.caltech.edu}
\author{Zhi-Yun Li}
\affil{Astronomy Department, University of Virginia, Charlottesville, VA 22903;
zl4h@protostar.astro.virginia.edu}
\author{Roger Blandford}
\affil{Theoretical Astrophysics, California Institute of Technology,
130--33 Caltech, Pasadena, CA 91125;
rdb@tapir.caltech.edu}

\begin{abstract}
We present time-dependent, numerical simulations of the
magneto-centrifugal model for jet formation,
in an axisymmetric geometry,
using a modification of the {\software Zeus3D}
code adapted to parallel computers.
The gas is supposed cold with negligible thermal pressure throughout.
The number of boundary conditions imposed on the disk surface is that
necessary and sufficient to
take into account
information propagating upstream from the
fast and Alfv\'en critical surfaces,
avoiding over-determination of the flow and unphysical effects, such as
numerical ``boundary layers'' that otherwise
isolate the disk from the flow and produce impulsive accelerations.

It is known that open magnetic field lines can either trap or propel
the gas, depending upon the inclination angle, $\theta$, of the
poloidal field to the disk normal.
This inclination is free to adjust, changing from trapping to
propelling when $\theta$ is larger than $\theta_c\sim 30\arcdeg$;
however, the ejected mass flux is imposed in these simulations as a
function of the radius alone.  As there is a region, near the origin,
where the inclination of field lines to the axis is too small to drive
a centrifugal wind, we inject a thin, axial jet, expected to form
electromagnetically near black holes in AGN and Galactic superluminal
sources.

Rapid acceleration and collimation of the flow is generally observed
when the disk field configuration is propelling.
We parametrize our runs using a magnetic flux
$\Psi\propto R^{-e_\Psi}$ and mass flux $j=\rho v_z\propto R^{-e_j}$.
We show in detail the steady state of a reference run with parameters
$e_\Psi=-1/2$, $e_j=3/2$, finding
that the wind leaves the computational volume in the axial direction with
an Alfv\'en number $M_A\sim 4$, poloidal speed
$v_p \sim 1.6v_{K0}$, collimated inside an angle
$\theta\sim 11\arcdeg$.
We show also the thrust $T$, energy $L$, torque $G$ and
mass discharge $\dot M$ of the outgoing wind, and
illustrate the dependence of these quantities with the exponents
$e_\Psi$ and $e_j$.
\end{abstract}

\section{Introduction}

Astrophysical jets are fast and well-collimated flows, observed in a
wide variety of astronomical systems, ranging from young stellar objects
(e.g.\ \cite{Lada}), Galactic superluminal sources
(e.g.\ \cite{Mirabel}, \cite{Hjellming}),
to quasars (e.g.\ \cite{Begelman84}).
A single mechanism may be responsible for their formation and collimation
(e.g.\ \cite{Livio97}).
A candidate model is the magneto-centrifugal
mechanism proposed by Blandford and Payne (1982),
investigated for instance in
\cite{Clarke86}, \cite{Lovelace86}, \cite{Pudritz86},
\cite{Konigl89}, \cite{Ostriker97}, and reviewed
recently by K\"onigl \& Pudritz (1999; see, however, \cite{Fiege96}
for an alternative model).
In this picture, parcels of cold gas are stripped from the surface of a
Keplerian disk and flung out along open magnetic field lines by
centrifugal force.  At large distances from the source region,
rotation winds the field lines up into concentric loops around the axis.
These magnetic loops pinch on the outflow (or wind) and collimate it
into a narrow jet.

The original wind solution of Blandford \& Payne (1982) is
axisymmetric, steady-state and self-similar.
The self-similarity
ansatz is expected to break down near the rotation axis and at large
distances, where appropriate boundary conditions must be imposed.
There have been several studies seeking non-self-similar steady-state
wind solutions.
Sakurai (1985, 1987) and \cite{NS94} took the approach of starting
directly from the time-independent MHD equations and found
steady-state solutions iteratively.  However, no general
solutions are available at the present from this approach, due to the
great complexity of the critical surfaces whose loci are unknown a
priori (\cite{Heinemann78}).  Moreover, there is no guarantee that a
steady-state
solution can be found at all for a given set of boundary conditions.
A time-dependent approach provides a
surer method for studying MHD outflows, both steady state or not.
This more flexible approach has been adopted by several authors
(e.g., \cite{Lind89};
\cite{SN93}; \cite{OuyedAPJ1}; \cite{Romanova97};
\cite{Ustyugova98}; \cite{KG99}; \cite{BT99}).
It is also the approach we shall take.

Following Stone \& Norman (1993) and
Ouyed \& Pudritz (1997), we treat the Keplerian
disk as a boundary and use the {\software Zeus} MHD code to simulate
the time-dependent, disk-wind.  The main subtleties come from treating
the disk-wind boundary --the disk surface-- and the region close to
the rotation axis.  A special treatment of the disk-wind boundary is
warranted, because this is the place where matter is
ejected into the wind
and where magnetic field lines are anchored.  On this
boundary, some of the flow variables can be fixed, but not all of them.
This point is illustrated most clearly by steady-state MHD winds,
which have three, well-known critical surfaces
---slow, Alfv\'enic and fast (\cite{Heinemann78}).
Crossing each of these critical surfaces imposes a set of constraints
on the flow variables.
Similarly, it is impossible to assign arbitrary fixed values
to all the quantities at the disk boundary: some quantities must
be left free to adjust according to flow conditions in the sub-fast region.
Failure to do so will overdetermine the problem, forming numerical
boundary layers with local discontinuities and impulsive acceleration
of numerical origin close to the disk, which
may cast doubts about the origin of the jet launched in the
simulation.
A numerical boundary layer may cause the effective boundary conditions
governing the wind flow
to differ from those imposed at the disk, as mentioned in
\cite{Meier97}.
This loss of control is particularly inconvenient when doing
parametrical studies.
In this paper we prevent such overdeterminations by properly
taking into account
the existence of information going upstream from the critical
surfaces.
Our main improvement over previous {\software Zeus}-based works is to
fix only the necessary number of boundary conditions at
the disk surface.
Romanova et al.\ (1997) and Ustyugova et al.\ (1999) have
implemented a similar method for their Godunov-type code independently.

In addition, we derive a physically motivated method for treating the
outflow near the rotation axis.  Centrifugal acceleration in the axial region
does not happen in the magneto-centrifugal picture because the
centrifugal force vanishes on the axis.  A gravitational infall is
expected along the axis, which could disrupt the outflow from the disk
through streaming instabilities.  To prevent such infall, some
previous authors (e.g., \cite{OuyedAPJ1}) assumed a high (thermal
and/or turbulent) pressure to counterbalance the gravity of the
central object.  We, on the other hand, assume the presence of a narrow jet
near the axis as explained in \S\ref{magnetomodel}.
Such a jet could be driven either electromagnetically by a rotating
black hole (\cite{BZ77}) in the context of active galactic nuclei
(AGN) or by interaction of a central star with a disk stellar wind
(e.g.\ \cite{Shu91}) in the context of young stellar objects.

In this paper we describe in detail our numerical approach and
present our first results using the code.
In \S\ref{physmodel}
we explain the physical model,
the numerical code is introduced in \S\ref{computer} and the
the boundary conditions in \S\ref{boundary}.
Our results are described in \S\ref{results}, and
conclusions in \S\ref{conclusions}.
Further features of the numerical approach are presented in an Appendix.

In forthcoming papers we
shall explore the effects of the size of the jet-launching disk,
the dependence of the mass flux $j$ with
the inclination angle $\theta$, and perform a 3D study.

\section{Physical Model}
\label{physmodel}
\subsection{Equations and notation}
We will use axisymmetric
cylindrical coordinates $(z,R,\phi)$, such that the launching surface
is at $z=0$ and the axis at $R=0$.  A subindex $p$ will denote
quantities in the poloidal plane $(z,R)$.
The notation is similar to that in
Blandford and Payne (1982) or \cite{Lovelace86}.

The adiabatic MHD equations used here are
\begin{eqnarray*}
\frac{\partial\rho}{\partial t}+\nabla\cdot\left(\rho\ve\right)&=&0\\
\rho\frac{\partial\ve}{\partial t}+\rho\left(\ve\cdot\nabla\right)\ve
&=&-\nabla p+\rho\nabla\Phi_g+{\mathbf j}\times\B/c\\
\frac{\partial\B}{\partial t}&=&\nabla\times\left(\ve\times\B\right)=
\nabla\times\E\\
\frac{\partial u}{\partial t}+\nabla\cdot\left(u\ve\right)&=&-p\nabla\cdot\ve\\
p&=&(\gamma-1)u
\end{eqnarray*}
where
\begin{eqnarray*}
\rho&=&\text{matter density}\\
\ve&=&\text{velocity flow field}\\
\B&=&\text{magnetic field}\\
{\mathbf j}&=&(c/4\pi)\nabla\times\B=\text{current density}\\
\E&\equiv&\ve\times\B=-c{\mathbf E}\\
\Phi_g&=&\text{gravitational potential}\\
p&=&\text{thermal pressure}\\
u&=&\text{internal energy density (per unit volume)}\\
\gamma&=&\text{adiabatic index}
\end{eqnarray*}
We use a smoothed gravity,
$\Phi_g=-{\Omega_0^2 r_g^3}/{\sqrt{r_g^2+z^2+R^2}}$,
with $\Omega_0$ and $r_g$ fixed parameters.  The product
$v_{K0}=\Omega_0 r_g$ gives the scale of Keplerian speeds.

Some quantities are conserved along the fieldlines in the steady state
(\cite{Mestel68}), becoming functions of the magnetic flux $\Psi$ alone.
Two such quantities are the relation between mass and magnetic flux
$k=4\pi\rho v_p/B_p$ and the conserved angular velocity
$\Omega=R^{-1}\left(v_\phi-B_\phi v_p/B_p\right)$, which at
$z=0$ has the equilibrium value
$\Omega=\sqrt{R^{-1}\partial\Phi_g/\partial R}\equiv v_K/R$.
In steady state, the velocity is related to the magnetic field by
$\ve=\frac{k\B}{4\pi\rho}+R\Omega{\hat\phi}$.
Conservation also applies to the specific angular
momentum $\ell=R\left(v_\phi-B_\phi/k\right)$ and the
specific energy $e=v^2/2+h+\Phi_g-\Omega R B_\phi/k$, where $h$ is the
enthalpy, of small importance in our cold winds.
Finally, the entropy $S$ is conserved by our equations,
but it is not relevant here.
We will check these conserved quantities as a measure of the
final approach to steady state.

\subsection{Magneto-centrifugal acceleration model}
\label{magnetomodel}
Consider poloidal magnetic fieldlines emerging from the surface
of an accretion disk.  In steady state, the flow will be along these lines,
due to the frozen-in property of ideal MHD flows.  We can picture one element
of gas flowing along a fieldline as a `bead' sliding on a rigid `wire'
(\cite{Henriksen71}).
Let that bead start from $z=0$, at a slow speed or at rest.
The wire is rotating at a constant angular speed $\Omega$.  This will
exert on the bead a centrifugal force outwards, competing against
the gravitational force pushing the bead inwards.
For a lever arm large enough, the centrifugal force wins this
competition, and the field line propels the flow.
This happens when the inclination $\theta$ of the poloidal
fieldlines from the axis is larger that a critical angle
$\theta_c$, allowing the flow to
accelerate centrifugally along the fieldline as if shot by a sling.
This critical angle is $\theta_c=30\arcdeg$ for Newtonian gravity and
Keplerian rotation, only slightly changed by our smoothed gravity.

The fieldlines twist in the toroidal direction due to the
differential rotation between the corona and the disk.  This generates
a toroidal component $B_\phi$, whose magnetic pressure tends to
collimate the flow, decreasing $\theta$ along a fieldline for
increasing height $z$.
In addition, the gradient in the poloidal field pressure causes
focusing, depending on the shape of the flux function.
Critical surfaces are the Alfv\'enic
($v_p=B_p/\sqrt{4\pi\rho}\equiv v_{Ap}$) and fast surface
($v_p=B/\sqrt{4\pi\rho}\equiv v_{At}$).
The slow magnetosonic surface is not relevant
here due to our low sound speed.
Acceleration is due to the centrifugal and magnetic forces,
which taken per unit mass and projected along the poloidal fieldlines, are
$f_C=(v_\phi^2/R)\sin\theta$ and
$f_M=\left({\mathbf {j}}\times\B\right)\lfrac{\B_p}{c\rho B_p}
=-(1/8\pi\rho R^2)(\B_p/B_p)\nabla(R B_\phi)^2$
(\cite{Ustyugova98}).
These two forces must counter the projected gravitational pull
$f_g=-(\B_p/B_p)\nabla\Phi_g$.
Effective acceleration requires that the
collimation of the fieldlines is not complete before
the Alfv\'en surface is reached: otherwise the projected
centrifugal force would be too small.

Close to the axis, the fieldlines have an inclination lower than
the critical angle $\theta_c=30\arcdeg$, so they cannot propel the flow
centrifugally.  For the purpose of designing
this fully cold simulation, we cannot prevent gravitational infall by
the use of thermal pressure, which is an isotropic mechanism that
would push across fieldlines as well as along them, making a large
fraction of our simulation hot.
We assume instead a ballistic axial injection of matter close to
escape speed.  Its ram pressure provides a simple non-isotropic
mechanism, which achieves the numerical objective of stopping
gravitational backflow and isolating the region where the inclination
of fieldlines is too small, without compromising the footpoints of the
fieldlines with transverse flows, as a
pressure-driven mechanism might tend to do.
This jet core made of axial flow is not only a convenient numerical
device; it is physically
justified in AGN, and galactic superluminal sources, which
may create a central electromagnetic jet carrying Poynting flux and
exerting an anisotropic Maxwell stress at its surface.
(\cite{BZ77}).  The ballistic injection
thus represents the effects of a feature of the flow not covered by
the magneto-centrifugal model alone.

\section{Numerical Methods}

\subsection{Computer code}
\label{computer}
To perform our simulations we wrote a variant, called
{\software Zeus36}, based on the {\software Zeus3D} code, released by LCA
(\cite{ZEUS}; \cite{LCA}; \cite{SN92}).
{\software Zeus3D}
is an explicit, finite-differencing algorithm, running on an
Eulerian grid.  Conservation of mass and momentum components
is guaranteed by the code.
Magnetic fields are evolved using the constrained transport method
(\cite{EH}), which guarantees that no monopoles will be
generated, limited only by the numerical precision of the machine.
An advanced method of characteristics (\cite{HS})
is used to calculate $\E=\ve\times\B$ and the transverse Lorentz force.
A large base of tests has already been done on this community code,
and we have not changed the central difference scheme.

{\software Zeus36} is a modification of {\software Zeus3D}
written for parallel computers, such
as for instance the Intel Paragon and the Touchstone Delta.
It uses MPI --a standard message passing interface for parallel machines,
including networks of workstations.
This makes
{\software Zeus36} easily portable in the parallel computing world.
The 3D parallelization scheme is straightforward.
We subdivide the computational cubic grid into N smaller grids,
each given to one of the N compute nodes.  Communication between
the nodes is realized
by taking advantage of the boundary condition subroutines,
appropriately modified to allow communication between adjacent compute
nodes.
{\software Zeus36}
also improves upon the latest released versions of
{\software Zeus3D} (version 3.4) by
treating the boundary conditions at the corners and edges of the grid
more consistently.  Corners and edges are of less importance when the
code is not run in parallel.  They become more important in a parallel
treatment, where they
appear in the middle of the overall mesh rather than less obtrusively
at the boundaries of the calculation.
The user interface was modified to make it more
convenient to our usage on many different computer systems, serial and
parallel.
For instance, it was decided not to use the powerful precompiler
{\software Editor}
(\cite{EDITOR}), bundled into the {\software Zeus3D} release.
These changes, together with the use of dynamic memory allocation,
eliminate the need to recompile the main code when the grid size is changed.

{\software Zeus36} has been tested for both internal consistency and
consistency with {\software Zeus3D}.  We have made sure that the
output does not depend on the distribution or number of the nodes used
to compute a given grid.  To test consistency with {\software Zeus3D},
we have run some test problems using both programs, and compared their
output byte-by-byte.  The small departures observed were those due to
the improvement in the treatment of corners and edges: no difference
appeared after those were accounted for.

\subsection{Boundary conditions}
\label{boundary}
In these axisymmetric simulations the computational grid has four boundaries:
axis ($R=0$), outer radius ($R=R_{\max}$), disk ($z=0$) and outer height
($z=z_{\max}$).  The code enforces boundary conditions by assigning
values to the fields $\rho$, $u$, $\ve$, and $\E$ at the grid edge and
at a few ghost zones outside the computational grid.
The boundary conditions for $\B$ are enforced
indirectly from the values of $\E$, using the equation
$\lfrac{\partial\B}{\partial t}=\nabla\times\E$ to evolve $\B$
in both active and ghost zones, ensuring
$\nabla\cdot\B=0$ everywhere.
The exact location and grid staggering of these ghost zones follows
Stone \& Norman (1992).

\subsubsection{$R=0$}
The axis boundary is geometrically well-determined by
axisymmetry, and it has offered no complications either in implementing or
running.  The conditions are implemented as usual for the axis in
cylindrical coordinates.  In the calculation of ghost zone values for
$R<0$, scalar fields are reflected, and vector
components are reflected with a change of sign in the directions $R$
and $\phi$.

\subsubsection{$z=z_{\max}$}
The upper boundary is treated in our simulation using an outflow
boundary condition, as defined in the {\software Zeus3D} code.
Specifically, for $z>z_{\max}$
$A(z) = A(z_{\max})$ for all fields,
except for $v_z(z)=\max(v_z(z_{\max}),0)$.
This is not perfect, but, as the
physics of our simulation implies the existence of a supersonic,
super-Alfv\'enic accelerated outflow,
these imperfections are not
expected to have a large influence in the simulated flow.

\subsubsection{$R=R_{\max}$}
This is more delicate, especially at low
altitude where the flow is sub-Alfv\'enic,
and can influence the upstream flow.
In practice we use similar outflow conditions as for $z=z_{\max}$.
Label the fieldline that becomes trans-Alfv\'enic at $R=R_{\max}$, by
$\Psi=\Psi_1$ (Fig.~\ref{S1_mai}), and it is clear that artificial
conditions imposed at $R=R_{\max}$ for $\Psi>\Psi_1$ may induce
spurious collimation.
We must therefore explore the sensitivity of our results to our
treatment of this part of the flow (cf.\ \S\ref{engineering}).

In one example of an alternative prescription,
Romanova et al.\ (1997) have
used a ``force-free'' prescription, ${\mathbf j}_p||\B_p$,
that can be written as $\B_p\cdot\nabla(RB_\phi)=0$; a later proposal
{}from the same authors is taking
$\B_p\cdot\nabla(RB_\phi)=\alpha B_RB_\phi$, with $\alpha$ a constant
to be determined (\cite{Ustyugova98}), observing
that artificial collimation may occur for inappropriate box sizes.

\subsubsection{$z=0$}

\label{DiskBC}
Even more important is the treatment of $z=0$.
It represents the
launching surface of the model, located at the top of the disk, or the
base of the corona, above the slow magnetosonic surface.
Its properties will determine all the flow downstream.

The first question to address is the number of boundary conditions we are
allowed to fix at this surface.
This number is equal to the number of waves outgoing normally from the
boundary, equal to the number of characteristics.
We calculate it by counting the degrees of freedom of the system,
subtracting all constraints.
We have, in principle, seven degrees of freedom: the density, three
components of $\ve$ and $\B$, minus the constraint $\nabla\cdot\B=0$,
and also the internal energy, because
our simulation, while dynamically cold, keeps track of a small
internal energy.
Each crossing of a critical surface is also a constraint that
will remove one degree of freedom.
In principle there are three such surfaces, slow, Alfv\'enic
and fast; however,
the initial $\ve_p$ at $z=0$ is already larger than the slow
speed, so only two critical surfaces will be crossed.  We are left
with five independent waves out of the original seven.  Five boundary
conditions should be imposed.
A more detailed deduction of this result can be seen for instance in
\cite{Bogovalov97}.

We choose to fix the five fields $\rho$, $u$, $\E_\phi$, $\E_R$
and $v_z$ at the disk
--if launching were sub-slow, the presence of the slow critical
surface would reduce the number of independent
boundary conditions to the first four.
At $z=0$, the boundary conditions for
$\E_\phi$
and $\E_R$ are derived
{}from the infinite conductivity of the disk material, giving
$\E_\phi=0$ and $\E_R=R\Omega B_z$.
The condition for $\E_\phi=(\ve_p\times\B_p)\cdot\hat\phi$ implies
that $\ve_p||\B_p$, $v_R/B_R=v_z/B_z=v_p/B_p$.
The condition for $\E_R=v_\phi B_z-v_zB_\phi$ implies that
$v_\phi=R\Omega+B_\phi v_p/B_p=R\Omega+B_\phi v_z/B_z$.
Implementation is done by assigning values to
fields at the ghost zones $z\leq 0$.
Density, velocity and internal energy are defined at the ghost zones
as straightforward
functions of $R$, with $u=\rho c_{s0}^2/(\gamma-1)$ for a small
constant $c_{s0}$.
This fixes the mass flux $\rho v_z$; if our launching
were subsonic, this quantity should be determined from the crossing of
the slow critical surface.

The magnetic field evolves in time following
$\lfrac{\partial B_z}{\partial t}=R^{-1}\lfrac{\partial R\E_\phi}{\partial R}$,
$\lfrac{\partial B_R}{\partial t}=-\lfrac{\partial\E_\phi}{\partial z}$,
and
$\lfrac{\partial B_\phi}{\partial t}=
\lfrac{\partial\E_R}{\partial z}-\lfrac{\partial\E_z}{\partial R}$.
Fixing $\E_\phi=0$ at $z=0$ keeps $B_z(R)$ constant in that plane,
anchoring the fieldlines to the disk surface.  The angle
$\theta=\arctan(B_R/B_z)$
between the fieldlines and the vertical evolves
in the active zones; a similar evolution must be present in the ghost zones.
Otherwise, the fieldlines would present sharp kinks at the base of the wind
$z=0$, associated with large currents and localized forces which might mask
the launching mechanism.
An evolving $B_R$ requires an $\E_\phi(z)$ dependence in the ghost zones;
knowing that $\E_\phi$ vanishes at $z=0$, we choose $\E_\phi$
to be an odd function of $z$.
The resulting antisymmetric time dependence of the field $B_z$ and the
flux $\Psi$ allows $\theta$ to vary in approximately the same way in
both active and ghost zones, so that the final inclination of the
fieldlines is decided by the flow itself and its crossing of the
critical surfaces, producing a much better steady state than
alternative approaches in which the fieldline inclination is fixed.
Similarly we make $\E_R(z)$ symmetric around its known value at $z=0$,
defining $\E_R(z)=2R\Omega(R)B_z(R)-\E_R(-z)$ for $z<0$.
(An alternative possibility here would be extrapolating
the fieldlines inside the ghost zones;
however, this is somewhat more difficult to implement.)
With this choice, one term of the time dependence of $B_\phi$ is
even in $z$.  We make the other term also even, by requiring
$\E_z(z)=\E_z(-z)$, which allows this quantity to vary freely.
Finally we need values for $v_R$ and $v_\phi$.  We will take them from
$v_R=B_Rv_z/B_z$ and $v_\phi=R\Omega+B_\phi v_z/B_z$, which we
implement using the values of $B_z$ and $\Omega$ at $z=0$.
Following the variation of these
quantities inside the ghost zones is also a possible option.

In some simulations we find occasions
--early during the run, long before steady state is
approached-- in which $v_z<0$ at the first active zone close to the disk.
In that case we modify the boundary conditions, allowing the disk
to absorb the backflow,
preventing the numerical artifacts shown in the Appendix.
The ghost values at $z\leq 0$ for the fields $\rho$, $v_R$ and $v_\phi$
are taken temporarily from their values at the first active zone with $z>0$,
$v_z$ is set to zero and the internal energy to $u=\rho c_{s0}^2/(\gamma-1)$.
Thus backflow is absorbed, and will eventually disappear.

In this paper we artificially prescribe $j=\rho v_z$ as a function of
$R$ alone, although mass loading is effective only for
$\theta>30\arcdeg$.  However, in the steady state simulations
presented here this inequality is satisfied for most of the disk
outside the innermost region, justifying our choices after the fact.
We shall return to this in a following paper, in which $j$ will be
allowed to depend also on $\theta$.

The functions of radius are parametrized using a combination of
exponents and softening radii:
\begin{eqnarray*}
\rho(R)&=&\rho_0/\sqrt{(1+(R/r_\rho)^2)^{e_\rho}}\\
B_z(R)&=&B_{z0}/\sqrt{(1+(R/r_b)^2)^{e_b}}\\
\Psi(R)&=&
\left\{
\begin{array}{l}
2\pi(B_{z0}r_b^2)\frac{\left(1+(R/r_b)^2\right)^{1-e_b/2}-1}{2-e_b}\\
\pi(B_{z0}r_b^2)\log\left( 1+(R/r_b)^2\right)\ \ \text{if $e_b=2$}
\end{array}
\right.\\
v_z(R)&=&
(v_{z{\text {inner}}}^\nu+v_{z{\text {outer}}}^\nu)^{1/\nu}\\
v_{z{\text {inner}}}&=&
\frac{v_{z0}/
\sqrt{(1+(R/r_{vi})^2)^{e_{vi}}}}{\sqrt{(1+(R/r_{vo})^2)^{e_{vo}}}}\\
v_{z{\text {outer}}}&=&
\left(f_{vo} R\Omega(R)\right)/\sqrt{(1+(R/r_{vo})^2)^{e_{vo}}}\\
\end{eqnarray*}
We have used $\nu=1$ (linear) and $\nu=2$ (quadratic).
For large $R$, $v_z\propto R^{(e_{vo}+1/2)}\equiv R^{-e_v}$,
$j\equiv\rho v_z\propto R^{-(e_v+e_\rho)}\equiv R^{-e_j}$, and
$\Psi\propto R^{-(e_b-2)}\equiv R^{-e_\Psi}$.

This implementation of disk boundary conditions uses some values of the fields
in the corona; there is
upstream propagation of information.  The velocity at $z=0$ is supersonic
but sub-Alfv\'enic; upstream propagation of waves is physically expected.
A simulation that omits this effect is incomplete,
and our results are, manifestly, sensitive to the treatment
of the surface conditions.
There is in principle a risk of numerical instability involved in
using this information in the code; fortunately,
our runs did not show this kind of instability.

\subsection{Initial conditions}
Naturally, we are interested in flows that are independent of our
starting conditions.
It turns out that transients generally decay in a few Alfv\'en
crossing times, and do not influence the late final flow.
Nevertheless, they deserve some detailed explanation.

Initially $\B_p$ is defined from
its boundary condition value
$B_z(R)$ at $z=0$, and
the initial position of the footpoint $(z=0, R=R_0)$ of a generic
poloidal fieldline passing through a point $(z,R)$.
This footpoint $R_0$ is found from the equation
$R_0^2-\left(R-m_\ell r_\ell (1+z/r_\ell)^{e_\ell}\right)R_0
-R m_\ell r_\ell=0$, where $B_{z0}$, $r_b$, $e_b$,
$r_\ell$, $e_\ell$ and $m_\ell$ are constant parameters.  The lengthscale
$r_\ell$ determines the location of the
initial footpoint $R_c$ of the critical fieldline
$\Psi_c$ with $\theta=\theta_c$, $e_\ell$ gives the initial curvature
and collimation of the fieldlines, and $m_\ell$ sets the maximum slope
$\tan\theta$ allowed for the fieldlines, through the prescription
$\tan\theta=e_\ell R/(r_\ell+R/m_\ell)$ valid at $z=0$.
If $m_\ell$ is set to infinity, all
angles $\theta$ are allowed, and $R_0(z,R)=R/(1+z/r_\ell)^{e_\ell}$.

The initial density at $z>0$ is given by
$\rho(z,R)=\eta_\rho\rho(z=0,R)$, with $\eta_\rho<1$.
Initial $v_z$ in the corona is defined similarly through
$v_z(z,R)=\eta_v v_z(z=0,R)$.
The initial value for the internal energy is
$u=\rho c_{s0}^2/(\gamma-1)$.

In all but one
of our simulations we have set the initial value of $B_\phi$ to zero,
and then consistently made the initial
$v_\phi(z=0,R)=v_K=R\Omega(R)$.  For simplicity,
this initial $v_\phi$ is chosen as independent of $z$.
Finally, the radial component of the velocity,
$v_R$, is set to zero in the bulk of the flow,
and to its boundary condition value $v_R=B_R (v_z/B_z)$ for
$z\leq z_{\min}$, where $z_{\min}$ is the value of $z$
at the first zone above the disk.

\section{Results}
\label{results}

\subsection{The reference simulation}
\label{steady state}
We first describe a fiducial or reference simulation which we shall
use to describe the various dependencies in our results.
In this simulation, we reach the steady state solution shown in
Fig.~\ref{S1_mai}.  The simulation box has $256\times 128$ active
pixels, with $0\le z\le 80.0$ and $0\le R\le 40.0$.
Initial and boundary conditions used to produce this run are shown
in Fig.~\ref{S1_boundary}.
The boundary density at $z=0$ is constant and uniform, $\rho=\rho_0=1$, with
initial $\rho=0.1\rho_0$ for $z>0$.
The Keplerian velocity scale is given by
$v_{K0}=\Omega_0 r_g=1$, with $r_g=\sqrt{3}$.
The function $v_z(z=0,R)$ is
determined by $v_{z0}=1.7$, $f_{vo}=0.1$, $r_{vi}=r_{vo}=r_g$,
$e_{vi}=2$, $e_{vo}=1$, making $e_v=1.5$.
Initially $v_z(z>0,R)=10^{-6}v_z(R)$, $B_\phi=0$ and
$v_\phi=R\Omega(R)$, independent of $z$.
The parameters determining $\B_p$ are
$B_{z0}=4 v_{K0} \sqrt{4\pi\rho_0}$, $r_b=r_g$, $e_b=3/2$,
$r_\ell=r_g$, $e_\ell=1$ and $m_\ell=\sqrt{3}=\tan60\arcdeg$.
The initial sound speed is set to $c_s=0.0002$, ensuring that pressure
is unimportant in this simulation.
The adiabatic index is $\gamma=5/3$.

The flow is then evolved until a steady state is reached.
Once in steady state,
the flow is accelerated up to $2.77$ times the Keplerian speed at the launching
point; the acceleration is
purely centrifugal and magnetic.  The maximum Alfv\'en number
$v_p/v_{Ap}=3.98$ is found close to the upper edge, where
also the maximum fast Alfv\'en number $v_p/v_{At}$ is found,
equal to $1.36$.  Collimation
of the fieldlines starts even before the flow crosses the
Alfv\'en surface $v_p=v_{Ap}$, as we can see in Fig.~\ref{S1_mai}.
The minimum Mach number is $140$, showing that pressure is
dynamically unimportant in this cold simulation.
The acceleration along the fieldline $\Psi=\Psi_e$ passing through the
outer edge of the grid is shown in Fig.~\ref{S1_speeds}.
Important fieldlines are $\Psi_1$, the last fieldline able to
make the Alfv\'en transition; $\Psi_c$, the fieldline making a
critical angle $\theta_c$ with the axis;
and $\Psi_0$, with footpoint at $R=r_{vi}$,
at the outer edge of the ``core'' inner injection, thus separating
this core from the wind region.

We have followed this simulation for a time equal to $2.6$ Keplerian
turns $2\pi/\Omega$ at $R=R_{\max}$
(and, equivalently, $94$
turns at the critical radius $R_c$, $170$ at the smoothing length
radius $r_g$).
The poloidal velocity and the magnetic field change by less
than $1.1\times10^{-4}$ in the last 10\% of the run.

The approach to steady state can also be observed in the figures
in the integrals of motion $\Omega$, $\ell$, $k$
and $e$, which are already functions of $\Psi$ alone, with the exception of
border effects for the smallest and largest values of $z$, important
mostly for the non-centrifugal core region, $\Psi<\Psi_0$.
We followed these integrals along all fieldlines in the wind $\Psi>\Psi_0$,
and found that on each fieldline the integrals depart from their averaged
value by less than 5\% for $\Omega$, 6\% for $\ell$, 7\% for
$k$ and 7\% for $e$ in the worst case.  This still overestimates
the errors; the corresponding standard deviations
divided by the mean are less than 0.3\% for
$\Omega$, and 3\% for the other integrals, further improved
to 0.9\% if the outermost region $\Psi>\Psi_1$ is excluded.

The Alfv\'en radius is defined by $R_A^2=\ell/\Omega$.
Our parameter choice is such that angular momentum is primarily
extracted from the disk by magnetic stress.
Therefore we can estimate $\ell$ for low values of $z$
by its magnetic term alone, and write an approximate lever-arm ratio as
$R_A/R_0\approx \sqrt{-(B_\phi/B_z) v_{Az}^2/v_zv_K}$.
Fig.~\ref{S1_leverarm} shows the observed value, the exact formula
and the estimate; all agree on a value around 3 for all fieldlines,
already large enough to justify approximating $\ell$ by its magnetic term.
In the formula of the estimate,
the boundary conditions determine all the quantities at $z=0$,
with the exception of the toroidal field $B_\phi$, which is allowed to
self-adjust in the simulation; assuming a value for the lever-arm
ratio gives an estimate for the toroidal field in steady state.

The collimation shifts
the position of fieldline $\Psi_c$ making a critical
angle with the disk surface.  In steady state $\theta_c$ is found at
$R_c=1.85r_\ell$, while initially it had been $R_c=1.63r_\ell$.
The integrals $\dot M(\Psi)$, $L(\Psi)$ and $G(\Psi)$ are the total
fluxes of mass, energy and angular momentum between the axis and a
fieldline $\Psi$.  They have very little numerical dependence on the height
$z$ of the surface used to perform the integration, because the
{\software Zeus} algorithm respects these conservation laws.
The thrust $T$ is defined here as the surface integral of
the kinetic energy term $\rho v_z^2/2$, calculated at $z=z_{\max}$.
The integrals $\dot M$ and $\Psi$ at $z=0$ are completely determined
{}from the boundary parameters, and have been done analytically.
Our results are shown in Table~\ref{tableS1}.

It is important to observe that in this steady state not only the fieldlines
and streamlines collimate, but also the density,
which is roughly a function of $R$ alone for large $z$, as shown in
Fig.~\ref{S1_density}, in spite of the boundary
condition at $z=0$, where $\rho$ is kept constant.
This result is in accord with previous asymptotic analysis
(\cite{Shu95}).

As a stringent quality check, we made contour plots of the
ratio between the projected magnetic and centrifugal
forces $f_M$ and $f_C$, defined in \S\ref{magnetomodel}.
Centrifugal acceleration dominates close to the disk, approximately up
to the point where $v_p$ achieves the local escape speed; magnetic
acceleration takes over afterwards.
There is no visible discontinuity associated with the
fieldline $\Psi=\Psi_1$;
this shows that the influence of the sub-Alfv\'enic
boundary at $R=R_{\max}$ is not unduly large.
However, we observe a triangle near the outer corner where
$f_M<0$, decelerating the flow --the vertices of this triangle are the
points $(z=60, R=40)$, $(z=80, R=30)$ and the outer corner.
To decide if this is a physical effect or an artifact caused by the edges,
we enlarged the box in both directions,
as shown below in \S\ref{superalfvenic}, finding that it
is indeed only an artifact.

\subsection{Parametric study}
\label{parameters}
We now explore the sensitivity of our solution to changes in the
functions
$v_z(R)$, $\rho(R)$ and $B_z(R)$, parametrized by the
exponents $e_v$, $e_\rho$ and $e_b$, or $e_j=e_v+e_\rho$ and
$e_\Psi=e_b-2$.

First we checked that $e_j$ is indeed a good parameter.
We did a variable density
simulation with the same value of $j=\rho v_z$ as the standard
run, but with $e_\rho=0.25$ instead of zero, and let it run for the
same physical time.  For $z>2$, the results of the variable density
simulation
coincide with the reference run; however, steady state is not
achieved at the outer corner (Figs.~\ref{N123lines} and \ref{N123den}).
Despite its slower convergence, it represents the same physical
steady state as the reference run, with the exception of very small
values of $z$, where the different profile of $\rho$ still has some
influence.
For a larger value of the density exponent, such as $e_\rho=1$,
the run fails to reach steady state in the same box as the
reference run.  However, simulations with a box size four times
larger (see \S\ref{superalfvenic}) show that the $e_\rho=1$ case is nearly
identical to the other cases.  The effects of the box size are
discussed in the next section.
We thank the referee for suggesting this test.

Having done this,
we next made a few simulations for different values of the parameters.
Fig.~\ref{Nvarious} shows the critical surfaces and fieldlines for
these runs.
Their shapes are qualitatively similar, with the remarkable exception
of the outer fieldlines for the flatter $j$ simulation
$e_j=0.5$, $e_b=1.5$; some of these lines
start pointing inwards instead of outwards. Not all fieldlines are
able to
propel the flow centrifugally at $z=0$; they must spend some initial
kinetic energy before acceleration can start.
Further tests can tell if this unexpected effect is physical or only a
numerical artifact.  Because the assumption $|\theta|>\theta_c$
is not fulfilled everywhere,
these tests should take into account that the mass
flux should be reduced for those fieldlines, as we will do
in future work by using a $j$ dependent on $\theta$.

Values of speed, Alfv\'en number, lever-arm ratio,
luminosity and torque are calculated at the point
$(z_{\max}, R_{\max})$, and given as functions of the exponents $e_b$,
$e_v$ and $e_\rho$ in Table~\ref{tableparameters}.
The first four colums are the simulations
represented in Fig.~\ref{Nvarious}.  The last
column is the variable density simulation shown in
Fig.~\ref{N123lines}, where we can already see that its flux at the
outer corner is slightly different from that of the reference run;
this is due to its insufficient convergence to the steady state in
that region of the computational volume.

\subsection{Testing of code convergence}
\label{engineering}
We want to test the convergence of the simulations to a steady
state solution by changing some computational and physical parameters
{}from our reference run.
\subsubsection{Resolution}
\label{lowres}
We reduced the resolution of the original standard run by one half,
using a box with $128\times 64$ active pixels,
$0\le z\le 80.0$ and $0\le R\le 40.0$.
Simulation converged to a steady state similar to the reference run,
with maximum $v_p/v_{Ap}=3.74$ and maximum $v_p/v_{At}=1.33$.
The resolution test is therefore passed.

\subsubsection{Box size: changes in both length and width together}
\label{superalfvenic}
Here we show a run done in a larger box with $256\times 128$ active pixels,
$0\le z\le 160.0$ and $0\le R\le 80.0$; this is the same
space resolution used in \S\ref{lowres},
with a box extending in space to twice the height and width.
Once a steady state was achieved, we checked that the lower quadrant of this
solution was compatible with the reference run.
For instance, in Fig.~\ref{S2_alfven} we compare the surfaces of constant
Alfv\'en number for the two runs with good results:
we recover the strongly accelerated flow found before.
The maximum Alfv\'en number in the lower quadrant of this run is 3.78,
and the maximum fast number is 1.34, which is close to the reference
run and even closer to the results of \S\ref{lowres}.
We also want to check the ratio of projected forces $f_M/f_C$,
especially near
the location of the outer corner of the reference run, now the
center of the computational box.
Looking into Fig.~\ref{S2_mc}, we see that $f_M/f_C$
in this region is smooth, with positive magnetic acceleration;
therefore we conclude that the deceleration observed in the reference run
was only an edge effect related to the outer corner.
As another indication that
our treatment of the outer boundary conditions
could be improved for $f_M$, this simulation
shows a tiny triangle of magnetic deceleration near its outer
corner.

\subsubsection{Box size: change in width only; a sub-Alfv\'enic case.}
\label{subalfvenic}
This run has also the same physical parameters, changing only the
shape of the computational box.
Here we cut the larger box from \S\ref{superalfvenic}
in half, obtaining a square
box with $128\times 128$ active pixels, $0\le z\le 80.0$
and $0\le R\le 80.0$.

The results show that the Alfv\'enic surface crosses the outer boundary
$z=z_{\max}$: the flow at the boundary $R=R_{\max}$ is fully sub-Alfv\'enic.
The last fieldline able to make the Alfv\'en transition,
$\Psi=\Psi_1$ crosses now the $z=z_{\max}$ edge.
It marks the separation
between two very different kinds of simulated flow.
Towards the axis, we have the usual magnetocentrifugally accelerated
flow in steady state.
But outside, in the sub-Alfv\'enic region,
we have a complex structure out of steady state,
with filaments of magneto-centrifugal
acceleration approximately parallel to the fieldlines, wide
islands where the projected magnetic force $f_M$ is negative,
and sharp variations in the magnetic twist $B_\phi/B_p$.
Collimation is excessive, slowing down the acceleration of the flow.
The total projected force along fieldlines, $f_g+f_M+f_C$, is negative
in a sizable part of the box, instead of only around the axis.

As we have shown in \S\ref{superalfvenic},
this complex flow disappears in a larger and axially elongated box;
it is made possible by the finite size, which sets artificial limits
on the critical surfaces.
The fieldlines with $\Psi<\Psi_1$ are able to reach the Alfv\'en surface,
carry the proper critical point information,
and propagate it downwards to the disk.
Due to the shape of the box here, most fieldlines have
$\Psi>\Psi_1$, with critical points at $z>z_{\max}$,
inaccessible to the simulation.
Mathematically, the assumptions used to decide the number of functions
to fix at the disk surface are wrong, because the critical points are
not reached.  In this region, each streamline
is almost independent from each other, creating the possibility of
filaments.

We want to check our assumption that the different behavior of our
square and non-square boxes is due to the location of the Alfv\'en
surface.
To do that, we moved the Alfv\'en surface used in \S\ref{subalfvenic} by
reducing the intensity of the magnetic fields by half without changing
the box.
It allows the flow to become super-Alfv\'enic
inside the square computational box.  The flow is smooth,
showing again a case where
the difference between filamentary and smooth runs lays in the
relative position of the simulated Alfv\'en surface and the
computational volume.
We have also checked that the conserved quantities are
functions of $\Psi$.

\subsubsection{Large initial $B_\phi$: dependence on initial
conditions in the sub-Alfv\'enic case}
The setup used here is similar to the reference run.
It differs only in the initial values of $B_\phi$ and $v_\phi$, set by
$v_\phi(z=0,R)=(1-f_{B\phi})v_K$ and $B_\phi(z=0.R)=f_{B\phi}v_K B_z/v_z$,
with $f_{B\phi}$ a small fraction, equal to 0.01.
For simplicity, we have kept using initial values of $B_\phi$ and
$v_\phi$ independent of $z$; we could have chosen them to make
the initial $\Omega$ a function of $\Psi$, but the initial values for
$B_\phi$ would still be away from steady state.

Here we have a large initial
toroidal field, which provides excessive collimation of the fieldlines,
making the angle $\theta$ too small for magnetocentrifugal acceleration.
Also, the magnetic force along the fieldlines is such that
${\cal B}=(R B_\phi)^2$ acts like a pressure;
$\cal B$ grows along the fieldlines, stopping the acceleration of the flow.
The results present filamentation, low acceleration, and high twist.
This flow is sub-Alfv\'enic in all the volume.
We observe here that initial conditions can matter in these simulations,
provided that the Alfv\'en surface is not reached.

\subsubsection{Results of the convergence tests}
These tests tell that smooth acceleration of the simulated flow
is obtained when a majority of the fieldlines reach the critical
Alfv\'en surface inside the computational volume.
If this condition is met, the simulation results depend mostly on
the disk boundary conditions, with little dependence on the outer
boundary conditions.  The transients are quickly
swept up by the advance of the fast accelerating flow.
The opposite happens when
too many fieldlines lay in the $\Psi>\Psi_1$ region:
then the runs show filamentation, excessive collimation,
dependence on initial and outer boundary conditions,
and --obviously-- a slow acceleration.
Controlling the position of the critical
surfaces inside the computational box can improve convergence,
which will help in the setup of future simulations.
We plan to utilize this idea in a future work, in which we will
set up initial and boundary conditions
such that the rotating disk will be enclosed by an Alfv\'en surface,
fully nested inside the computational box; this setup should
minimize the influence of the outer boundaries.
As shown in \S~4.3.3, the shape and size of the simulation
box can sometimes introduce artificial nonsteadiness to a
sub-Alfv{\'e}nic flow.
Caution must be observed also in steady-state sub-Alfv\'enic
flows, due to their possible dependence on the outer boundary.

The tendencies to filamentation, both of numerical and physical origin,
show that the stability and interaction between the various
streamlines in this flow are not fully explored yet.
This encourages us even more to follow up
this work with a 3D simulation, for which this work will be the
starting condition.  The parallel code necessary to do such
massive computation is already written and tested.

\section{Conclusions}
\label{conclusions}
This paper is the first in a series exploring various aspects
of jets launched from accretion disks.
Here we have presented the code
and shown the results of cold, axisymmetric, time-dependent
simulations,
paying special attention to the treatment of the
jet-launching surface and the region near the rotation axis.
{}From these simulations we conclude that:

1)\ As expected, cold, steady jets can be launched smoothly from Keplerian
disks by the magneto-centrifugal mechanism.  The same mechanism is
also effective in collimating the outflows.  Collimation is observed
both in the shape of the fieldlines and in the density profiles, which
become cylindrical and thus jet-like at large distances along the
rotation axis, in agreement with asymptotic analysis.

2)\ The magneto-centrifugal mechanism for jet production is robust to
changes in the conditions at the base of the jets, whose consequences
are consistent with expectations.

3)\ Steady-state jets obtained in earlier studies using an essentially
incomplete treatment of the disk boundary conditions are qualitatively
unchanged when this deficiency is rectified.  It remains to be shown,
however, whether the same is true for non-steady jets.

4)\ The structure of the outflows is insensitive to the initial density
or flow speed at the injection surface individually, as long as their
product (i.e., the mass flux) is kept constant.

5)\ The size of the simulation box can strongly influence the outcome of
a simulation unless most field lines pass through the Alfv\'en surface
within the box.  As a result, simulations that produce mainly sub-Alfv\'enic
disk outflows, steady or not, must be treated with caution.
This highlights the pressing
need for box-invariant simulations, which are under way.

6)\ {\software Zeus3D} can be parallelized so that solutions from
different subgrids match smoothly.

\acknowledgments{
This research was performed in part using the
CACR parallel computer system operated by Caltech.
We acknowledge useful discussions with James Stone, John Hawley,
David Meier and Shinji Koide.
This research was supported by NSF grant AST-95 29170 and NASA grant 5-2837.
RB thanks John Bahcall for hospitality at the Institute for Advanced
Study and the Beverly and Raymond Sackler Foundation for support at the
Institute of Astronomy.
}

\appendix
\section{Two numerical artifacts related to backflow}
\label{artifacts}
\label{backflow}
We mention here some numerical problems that had to be overcome in
developing the code, in the spirit of helping developers of similar codes.
A simulation can develop backflow through at least two physical mechanisms.
One is simple gravitational infall, which operates mostly in the axial region,
where the magneto-centrifugal acceleration fails.
A second mechanism is cocoon backflow.  Due to the differential
Keplerian rotation, the inner parts of the flow launch a jet earlier
than the outer.  The cocoon of this jet may produce a temporary backflow.
When the backflow impinges into the disk boundary,
our boundary conditions shown in \S\ref{DiskBC} carefully
allow the disk to absorb it.  This easy prescription
avoids the problems discussed below.
However, in our earliest simulations,
our disk boundary was independent of the
sign of $v_z$ in the first active zone.
This created two kinds of trouble: false acceleration, and filamentation.

\subsection{False acceleration}
In this scenario,
backflowing material tries to get back towards the disk, where
it finds a boundary condition imposing a positive speed outwards.
Compression increases the density close to $z=0$.  This increased mass
is then flung out at the speed $v_z$ given by the boundary condition imposed.
No physical process is involved: only a badly chosen boundary.
This false acceleration is seen more clearly when we reduce the
axial injection, allowing the gravitational infall to go all the way
into the disk; with incorrect boundary conditions,
it quickly bounces back unphysically.

\subsection{Non-physical filamentation}
Let's suppose now that, early in the simulation,
a portion of the cocoon backflow touches the disk
at some radius $R=R_1$.
If the disk boundary condition
insists on imposing a forwards speed $v_z>0$ to the material,
we will get an increase in density due to the artificial
compression of the coronal material.  If the ejection speed $v_z$
is not large,
a fast bounce-back will be avoided, and we will not get a
false acceleration like before.
Nevertheless, we still get a localized region, close to the disk,
where the density is abnormally large.  This larger density will be
associated with large localized $B_\phi$, and it will have a
large inertia; the magneto-centrifugal acceleration will not be
enough to lift the material up.  The region around $R_1$
will be a hole in the acceleration front; the jet will proceed
forwards for both smaller and larger radii, but it will stagnate here.
A succession of such holes would make the simulation look filamentary;
again, allowing the disk to absorb the backflow avoids this numerical effect.

\newpage

\figcaption{Reference simulation.\\
The local Alfv\'en number $v_p/v_{Ap}$ is shown in grayscale.
In thin black, fieldlines and vectors of ${\ve}_p$.
In medium black, the critical surfaces $v_p=v_{Ap}$, $v_p=v_{At}$
and the local escape speed line $v_p=\sqrt{-2\Phi_g}\equiv v_{\text{esc}}$;
the line $B_p=B_\phi$ is shown in dashes.
In thick black, some selected fieldlines: $\Psi_1$ passing through the
intersection between the Alfv\'en surface and the outer boundary;
$\Psi_0$ at the outer edge of the axial injection zone, separating the
injected core from the main wind;
$\Psi_c$, the fieldline whose inclination is critical at $z=0$;
and $\Psi_e$, passing through the outer edge of the grid.
\label{S1_mai}}
\includegraphics{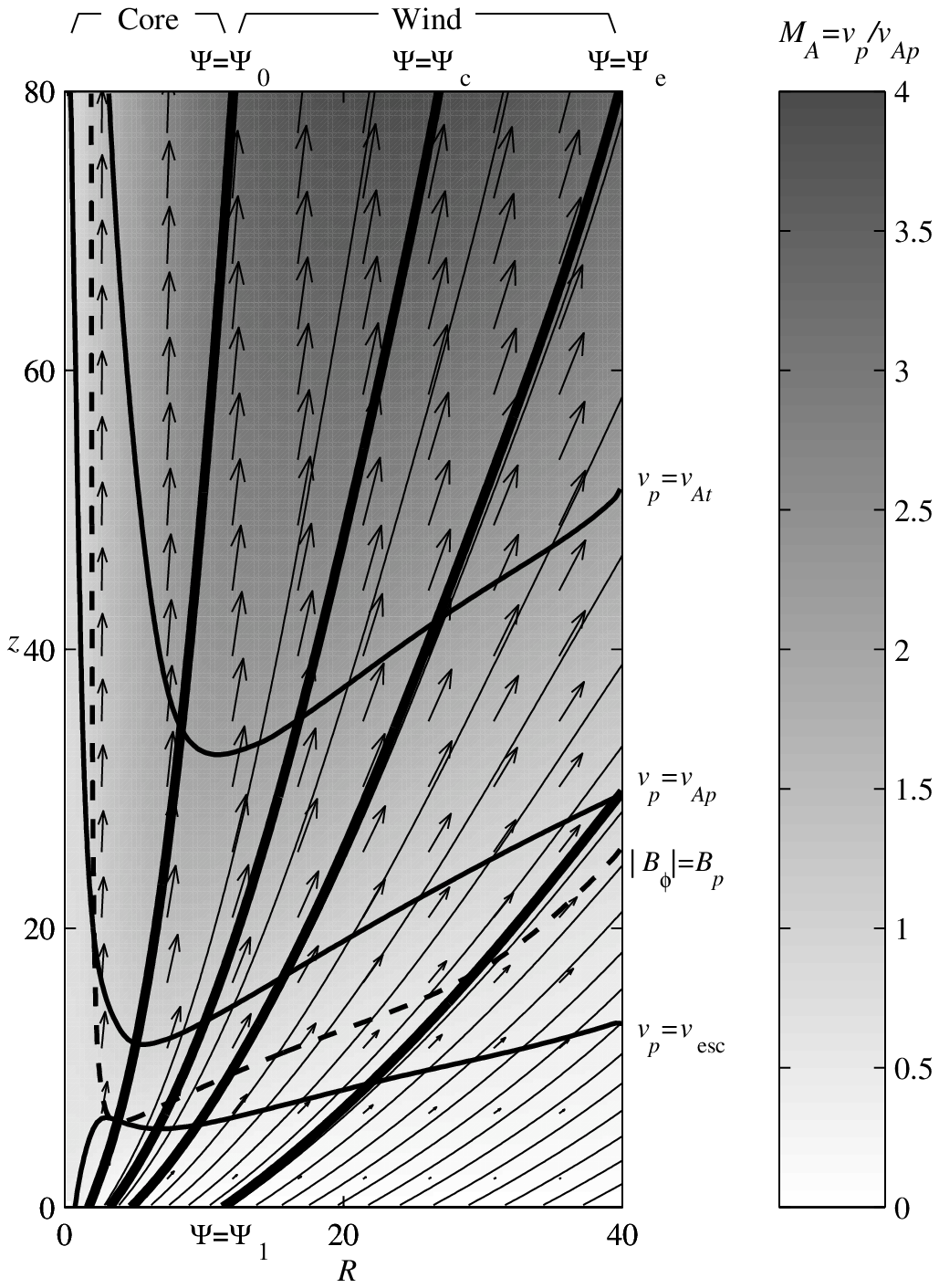}

\newpage
\figcaption
{Reference simulation.\\
Boundary and initial conditions for
$v_z$,
$v_{Az}=B_z/\sqrt{4\pi\rho}$,
$v_{Ap}$, $v_K\equiv R \Omega$, $B_z$ and
$\sqrt{-\Phi_g}=v_{\mathrm esc}/\sqrt{2}$
at $z=0$ as functions of $R$, normalized by the Keplerian speed scale
$v_{K0}=\Omega_0 r_g$.
\label{S1_boundary}}
\includegraphics{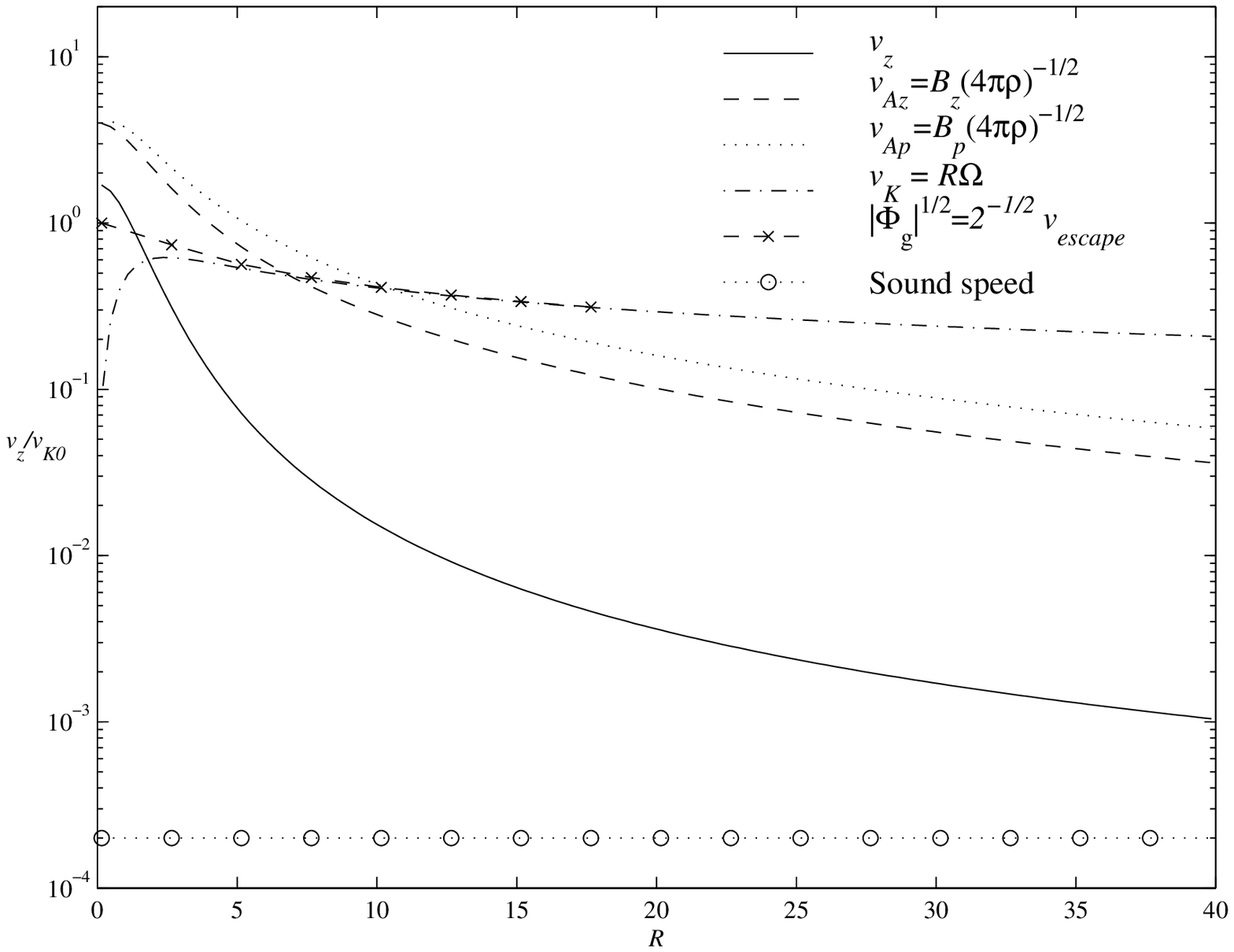}

\newpage
\figcaption
{Reference simulation.\\
Velocities along the fieldline $\Psi=\Psi_e$ passing through the outer
edge of the grid.
Plot of $v_p$, $v_{Ap}$, $v_{At}$, $v_{\text{esc}}$, $R\Omega$,
$R_0\Omega$ and $v_\phi$ vs.\ $z$, showing the growth of poloidal speed.
\label{S1_speeds}}
\includegraphics{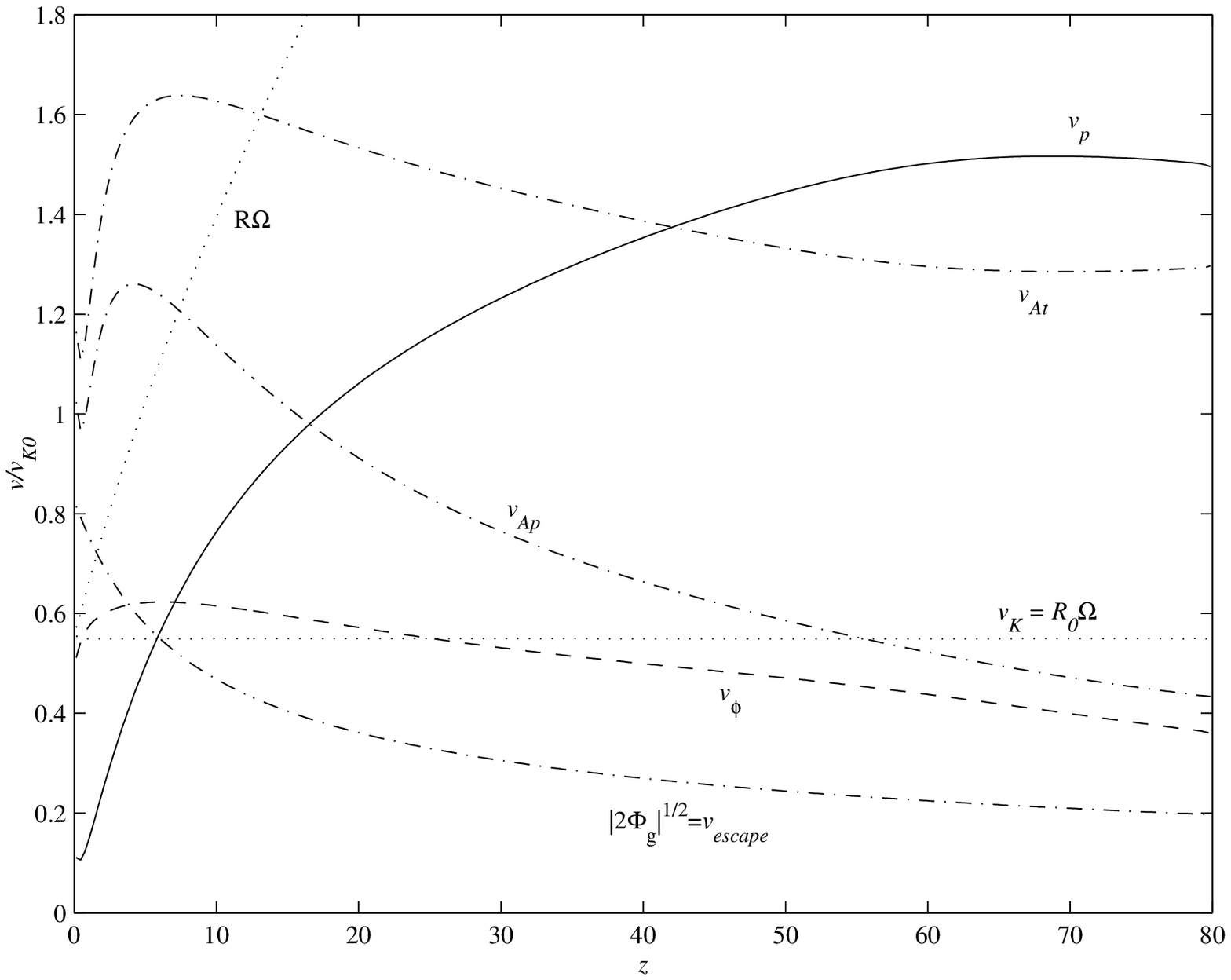}

\newpage
\figcaption
{Lever arm ratio $R_A/R_0$ (full lines),
compared to $\sqrt{\ell/\Omega}$ in dashes
and $\sqrt{-(B_\phi/B_z) v_{Az}^2/v_zv_K}$ in dots.
Both the flux $\Psi$ and the footpoint
$R_0$ are used to label the fieldlines.
\label{S1_leverarm}}
\includegraphics{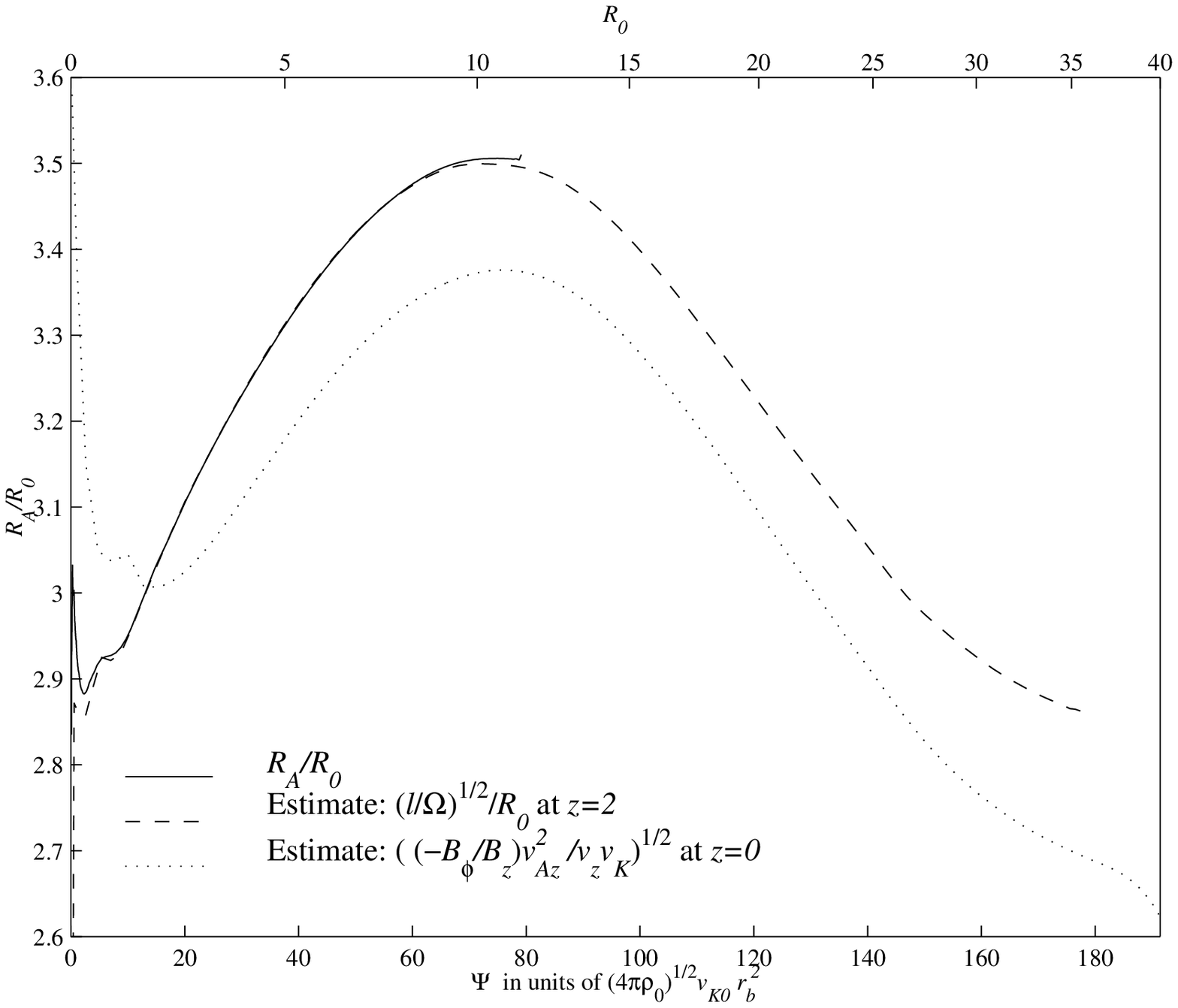}

\newpage
\figcaption
{Contours and greyscale of density in the reference run, showing collimation.
\label{S1_density}}
\includegraphics{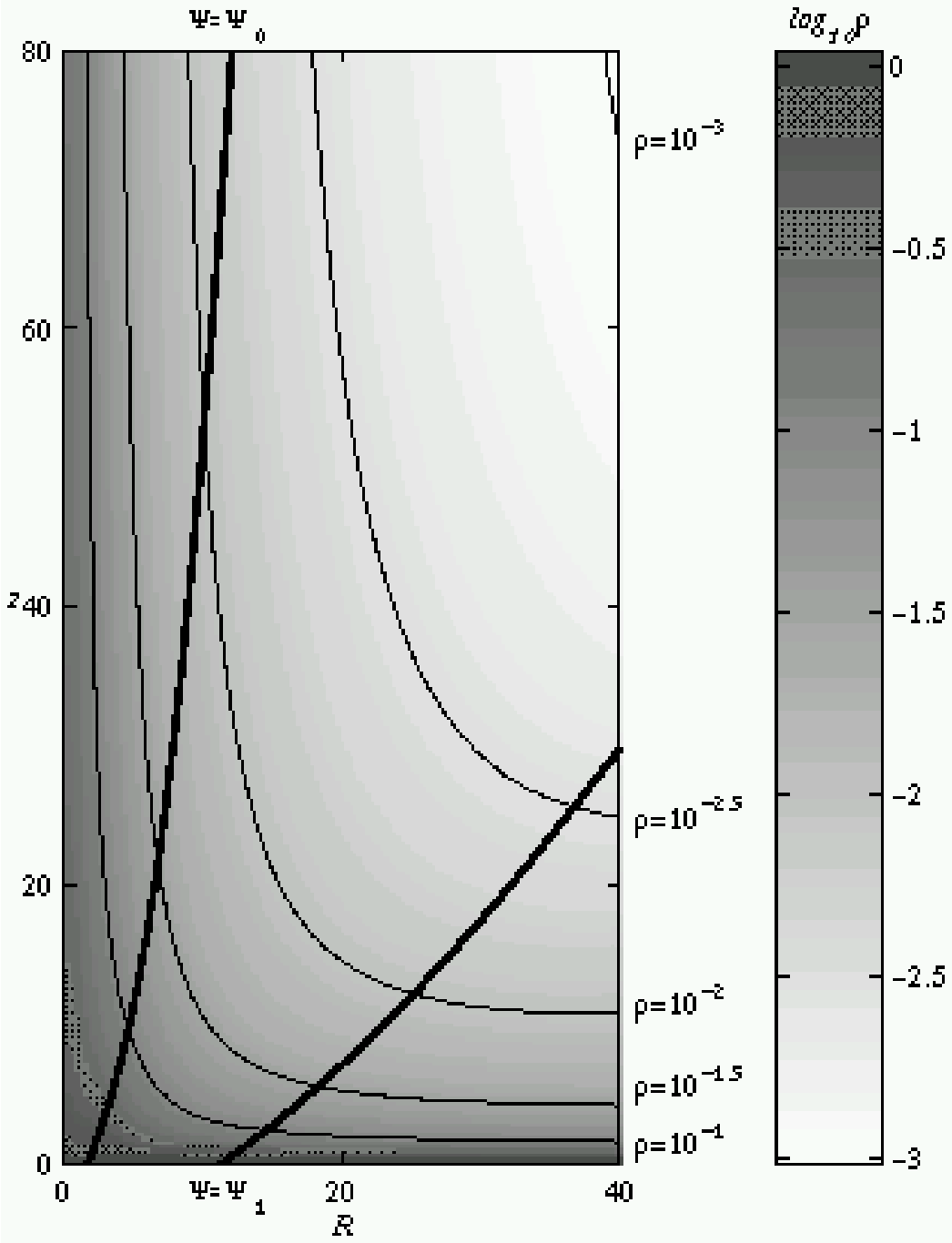}

\newpage
\figcaption
{A variable density simulation defined by
$e_\rho=0.25$, $e_j=1.5$, and $e_\Psi=-0.5$,
is compared to the
reference run, which has the same mass flux $j=\rho v_z$,
and a flat density profile at $z=0$,
defined by $e_\rho=0.0$, $e_j=1.5$, and $e_\Psi=-0.5$.
Fieldlines and contours of $M_A$ in these two runs are shown.
\label{N123lines}}
\includegraphics{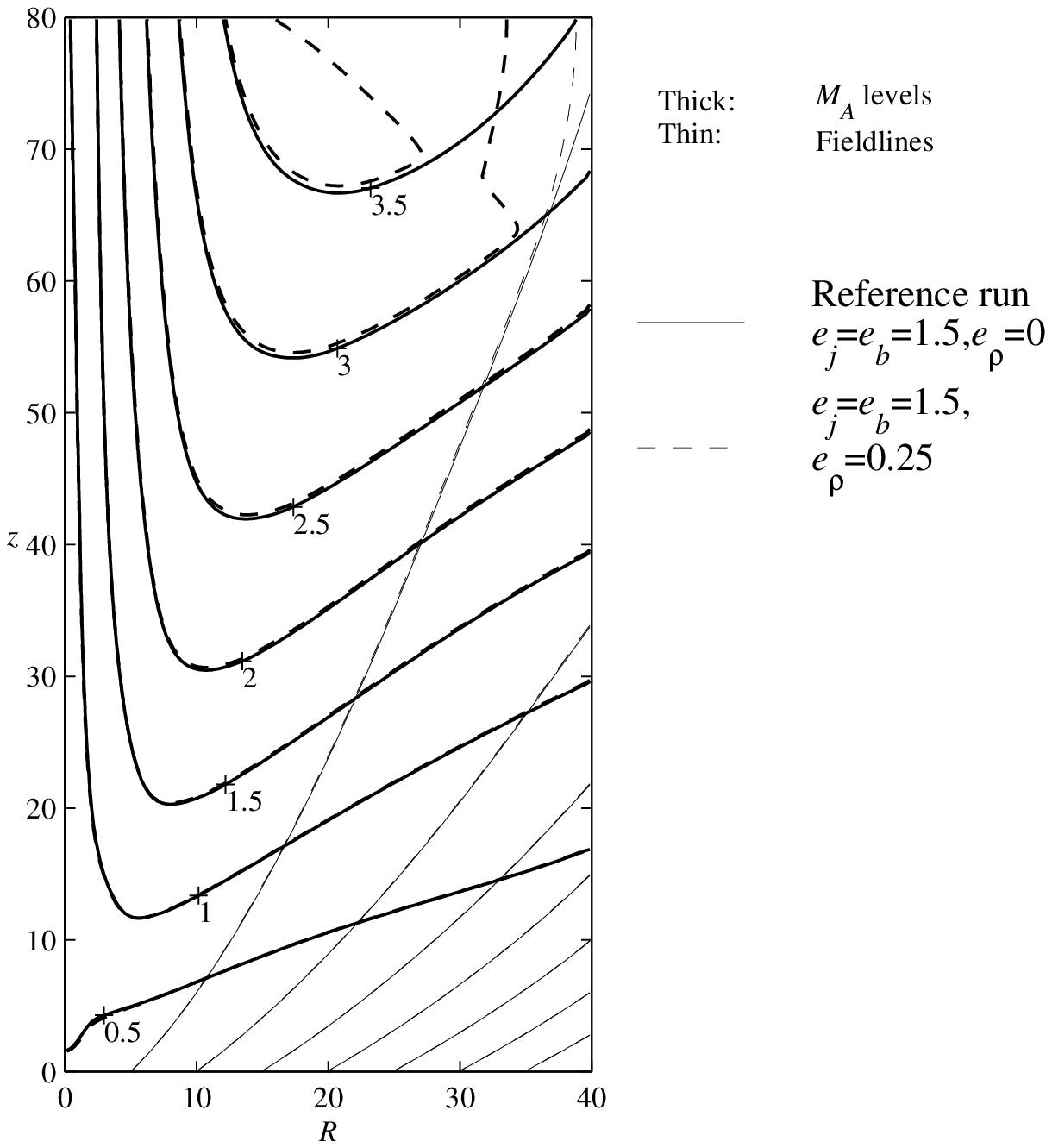}

\newpage
\figcaption
{The variable density simulation defined in Fig.~\ref{N123lines}
is again compared to the
reference run, this time
by showing contours of the density of one run divided by
that of the other.
\label{N123den}}
\includegraphics{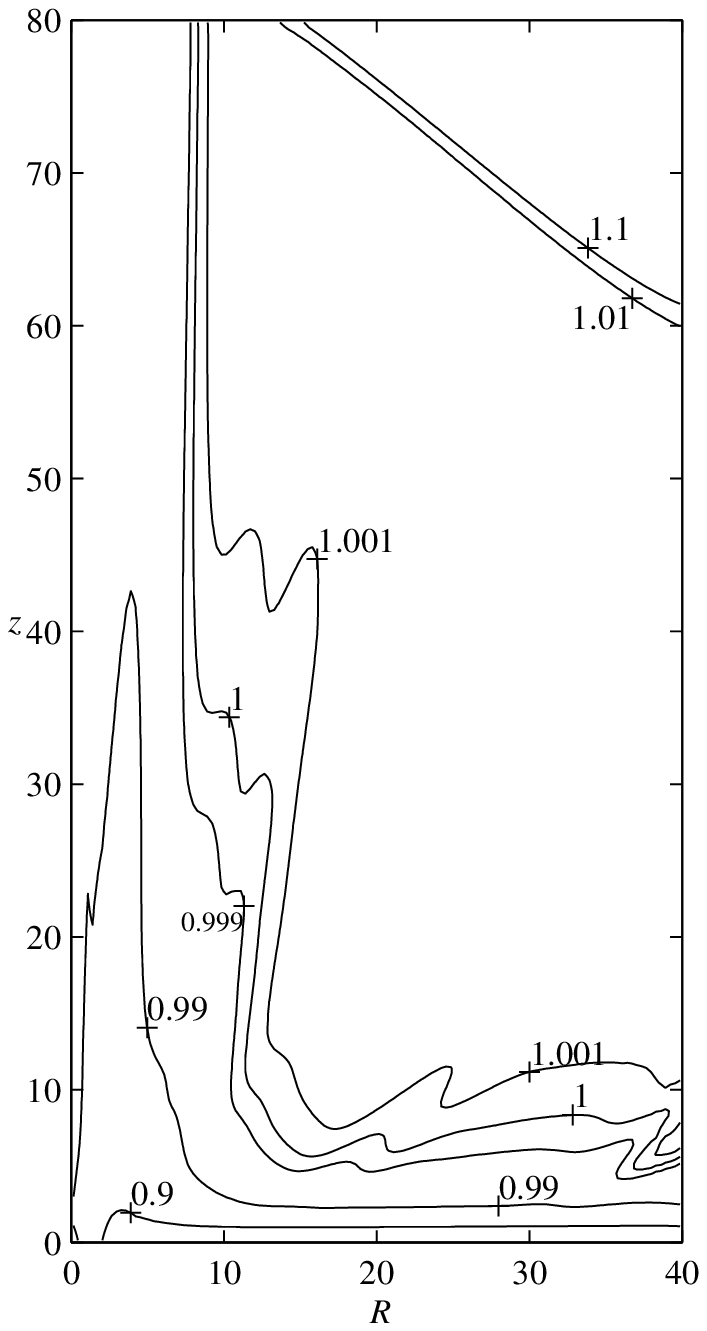}

\newpage
\figcaption
{Comparison of four runs.  Alfv\'en surface in thick lines, fast
surface in medium and fieldlines in thin.
Fieldlines have been chosen to pass through the same footpoints in the
four runs.
\label{Nvarious}}
\includegraphics{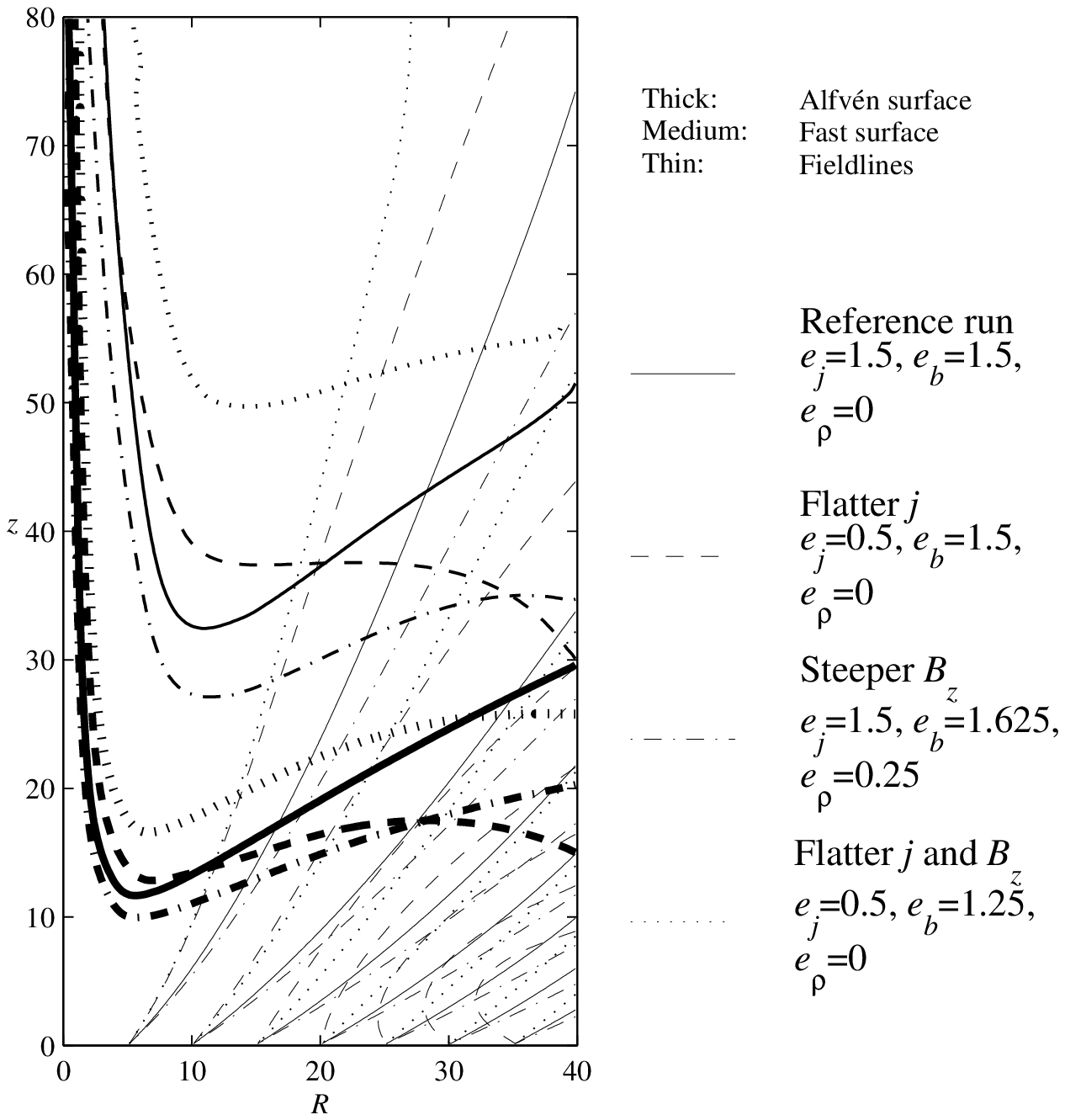}

\newpage
\figcaption
{Contour plots of $M_A=v_p/v_{Ap}$, for the reference run (solid
lines) and a larger and wider box (dashes), compared
to check convergence.
\label{S2_alfven}}
\includegraphics{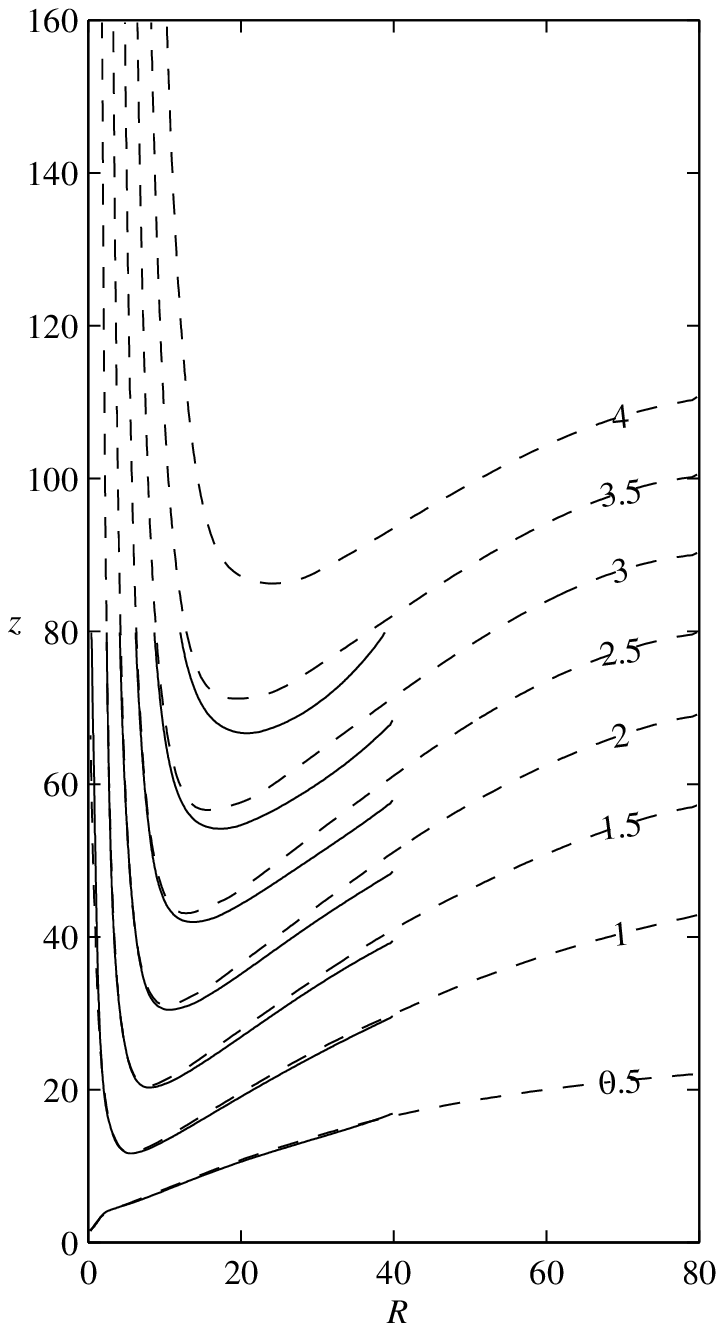}

\newpage
\figcaption
{Simulation on the larger and wider box.\\
In solid, the contour levels $f_M/f_C=0$, 1, 2, 3, and 4, show the
ratio of magnetic over centrifugal forces along
the fieldlines (shown in dots).
The $f_M=0$ level is a short, thick line at the outer corner.
\label{S2_mc}}
\includegraphics{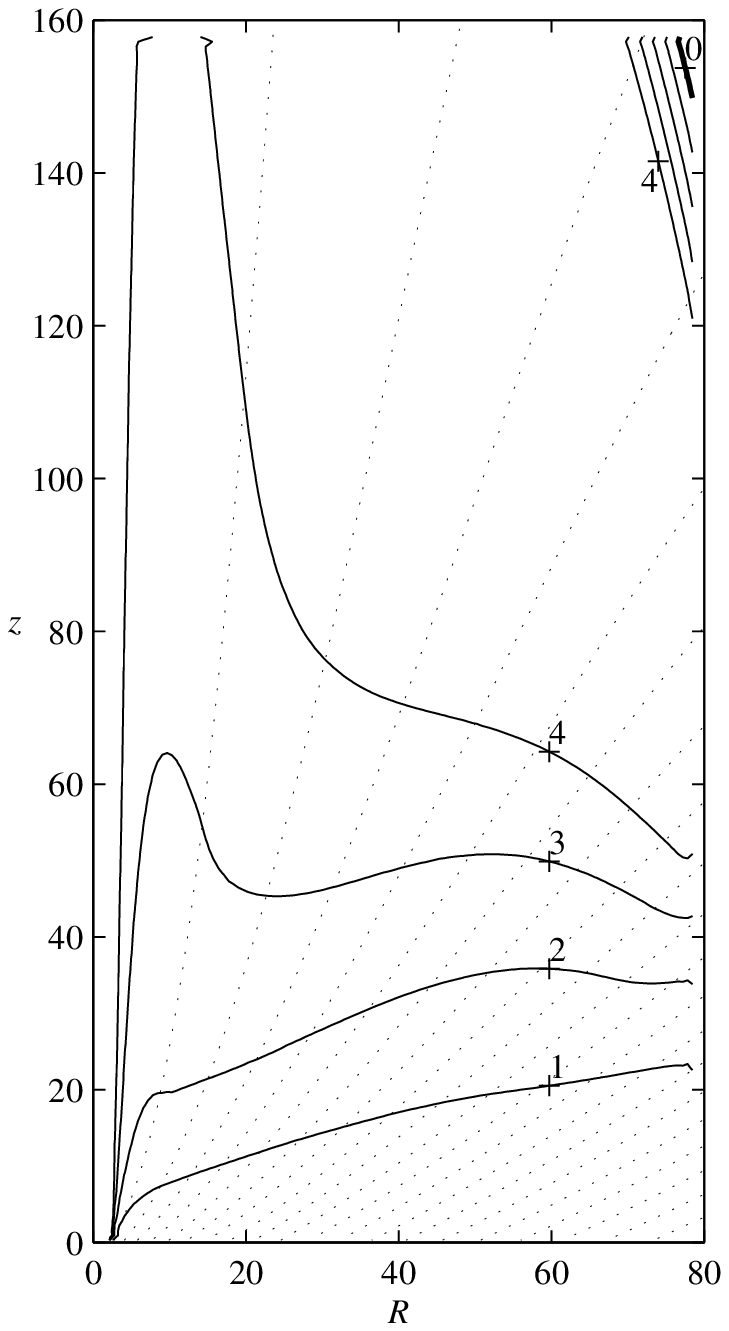}


\newpage
\begin{table}[p]
\caption{Mass discharge $\dot M$, luminosity $L$,
torque $G$, and thrust $T$ at selected fieldlines labeled by
their magnetic flux $\Psi$ or footpoint $R_0$.
$\dot M$ is in units of
$\rho_0v_{K0}r_b^2$, $L$ in $\rho_0v_{K0}^3r_b^2$,
$G$ in $\rho_0v_{K0}^2r_b^3$, $T$ in $\rho_0v_{K0}^2r_b^2$,
and $\Psi$ in units of $\sqrt{4\pi\rho_0}v_{K0}r_b^2$.
Thrust was calculated at $z=z_{\max}$.}
\label{tableS1}
\begin{tabular}{lccccccc}
   &
$\Psi_0$   & $\Psi_c$  & $\Psi_e$  & $2\Psi_c$ &
$3\Psi_c$ & $\Psi_1$  & $R_{\max}$ \\
\tableline
$L$    &
5.5&13.1&18.1&20.5&
24.0&25.0&28.5\\
$G$    &
8.8&31&54&71&
109&127&265\\
$\dot M$ &
 3.2& 6.0& 7.7&8.6&
10.2&10.8&14.8\\
$\Psi$ &
 9.5&22.6&35.8&45.2&
67.9&79.2&191\\
$R_0$   &
1.73&3.20&4.77&6.01&
9.41&11.35&40\\
$R_A/R_0$   &
2.94&3.14&3.29&3.38&
3.50&3.51&-- \\
$T$    &
1.9&4.1&5.5&--&
--&--&--
\end{tabular}
\end{table}

\begin{table}[p]
\caption{Dependence of some flow quantities on
the exponents $e_b$, $e_v$ and $e_\rho$.
Fluxes and thrust are calculated at the
outer corner of the grid, and adimensionalized as in Table~\ref{tableS1}.
The maxima of $v_p/v_{K0}$ are taken at $z=z_{\max}$.}
\label{tableparameters}
\begin{tabular}{lcccccc}
$e_v$              &   1.5   &   0.5   &   1.25    &   0.5   &   1.25    \\
$e_b$              &   1.5   &   1.5   &   1.625   &   1.25  &   1.5     \\
$e_\rho$           &   0.0   &   0.0   &   0.25    &   0.0   &   0.25    \\
\tableline
$e_j$              &   1.5   &   0.5   &   1.5     &   0.5   &   1.5     \\
$e_\Psi$           &  -0.5   &  -0.5   &  -0.375   &  -0.75  &  -0.5     \\
\tableline
$\Psi$             &  35.8   &  42.5   &   27.1    &   70.8  &   38.8    \\
$R_0$              &  4.77   &  5.64   &   3.91    &   7.68  &   5.15    \\
$R_A/R_0$          &  3.29   &  2.31   &   3.03    &   2.61  &   3.31    \\
Max.\ $M_A$        &  3.98   &  5.06   &   5.26    &   2.66  &   3.78    \\
Max.\ $v_p/v_{K0}$ &  1.67   &  1.33   &   1.67    &   1.31  &   1.54    \\
$\dot M$           &  7.73   &  15.02  &   6.89    &  19.54  &   8.05    \\
$L$                &  18.1   &  23.5   &   14.5    &  34.3   &   18.8    \\
$G$                &  54     &  85     &   38      &   162   &   58      \\
$T$                &  5.5    &  8.9    &   4.8     &   10.4  &   5.2     
\end{tabular}
\end{table}

\end{document}